\documentclass{article}
\usepackage[dblblindworkshop,final]{neurips_2026}
\usepackage[utf8]{inputenc} % allow utf-8 input
\usepackage[T1]{fontenc}    % use 8-bit T1 fonts
\usepackage{hyperref}       % hyperlinks
\usepackage{url}            % simple URL typesetting
\usepackage{booktabs}       % professional-quality tables
\usepackage{amsfonts}       % blackboard math symbols
\usepackage{nicefrac}       % compact symbols for 1/2, etc.
\usepackage{microtype}      % microtypography
\usepackage{xcolor}

\usepackage[dvipsnames]{xcolor}
\usepackage{multicol}
\usepackage{graphicx}
\usepackage{lipsum}
\usepackage{multirow}
\usepackage{comment}
\usepackage{tikz}
\usetikzlibrary{shapes.geometric,positioning,arrows.meta,calc}

\author{%
  Charles Fleming \\
  Cisco Research, USA \\
  \texttt{chflemin@cisco.com} \\ % Replace with actual email
  \And
  Luca Muscariello \\
  Cisco Outshift, France \\
  \texttt{lumuscar@cisco.com} \\ % Replace with actual email
  \And
  Vijoy Pandey \\
  Cisco Outshift, USA \\
  \texttt{vijoy@cisco.com} \\ % Replace with actual email
  \And
  Ramana Kompella \\
  Cisco Research, USA \\
  \texttt{rkompell@cisco.com} \\ % Replace with actual email
}

\begin{document}

%%
%% The "title" command has an optional parameter,
%% allowing the author to define a "short title" to be used in page headers.
\title{A Layered Protocol Architecture for the Internet of Agents}
\workshoptitle{AgenticOS}
\maketitle

\begin{abstract}
Large Language Models (LLMs) can learn domain-specific languages such as APIs
and tool interfaces, enabling agents that act through standardized protocols
like MCP. But LLM context windows cannot grow indefinitely, so scaling beyond
individual capacity requires agent collaboration -- the "Internet of Agents"
(IoA). Current network stacks (OSI, TCP/IP) were designed for data delivery,
not semantic coordination, so we propose two new layers: an \textbf{Agent
Communication Layer (L8)} that standardizes message envelopes, speech-act
performatives, and interaction patterns; and an \textbf{Agent Semantic Layer
(L9)} that negotiates a shared context, grounds and disambiguates content
against it, and provides consensus primitives for population-scale
coordination. On two real datasets and two LLMs, L9's schema-grounded
disambiguation improves accuracy on ambiguous requests by 13--72 percentage
points over blind guessing, at no added cost when disambiguation is not
needed; an ablation shows this gain is schema-specific for open-domain QA but
not for structured slot-filling, where the protocol's round-trip structure
alone already suffices.
\end{abstract}

%%
%% The code below is generated by the tool at http://dl.acm.org/ccs.cfm.
%% Please copy and paste the code instead of the example below.
%%
%%
%% Keywords. The author(s) should pick words that accurately describe
%% the work being presented. Separate the keywords with commas.
% \keywords{Do, Not, Us, This, Code, Put, the, Correct, Terms, for,
%    Your, Paper}
%% A "teaser" image appears between the author and affiliation
%% information and the body of the document, and typically spans the
%% page.

%\received{20 February 2007}
%\received[revised]{12 March 2009}
%\received[accepted]{5 June 2009}

%%
%% This command processes the author and affiliation and title
%% information and builds the first part of the formatted document.

\section{Introduction}
Large Language Models (LLMs) have demonstrated remarkable advances, including
the ability to learn domain-specific languages (DSLs) such as APIs and tool
interfaces, enabling agents that act through tool calling and standardized
protocols like the Model Context Protocol (MCP)~\cite{mcp2024}. However, LLMs
face a fundamental constraint: context windows cannot grow indefinitely,
limiting the complexity any single agent can solve alone. As in traditional
computing, where single processors hit physical limits, the answer is
distribution and collaboration---agents must work together to tackle problems
beyond individual capacity~\cite{chopra2025}.

Agent collaboration requires more than data exchange; agents must share
\textit{understanding}. Current network stacks (OSI, TCP/IP) were designed for
reliable data delivery~\cite{zimmermann1980}, not semantic coordination, and
this gap has produced a proliferation of ad-hoc, incompatible agent protocols.
HTTP is a precedent: it emerged ad hoc to fill an identical TCP/IP gap for
hypermedia retrieval, then was forced to accrete state and new verbs as the
web grew, before HTTP/2/3~\cite{thomson2022} finally gave it real transport
primitives and made it the internet's de facto ``narrow
waist''~\cite{popa2010}. Agents are at an identical inflection point; we argue
for formalizing the missing layers now rather than repeating two decades of
ad-hoc accretion.

\textbf{The Protocol Challenge:} Current protocols like A2A~\cite{a2a2024} and
MCP~\cite{mcp2024} standardize the \textit{syntax} of agent messages---envelopes,
performatives, interaction patterns---but cannot enforce \textit{semantic
agreement}. Two failure modes follow. First, a syntactically valid term (e.g.,
"switch") can carry ambiguous meaning that syntax alone cannot resolve. Second,
even an unambiguous message can be semantically \textit{underspecified}: a
perfectly parseable "book a ticket to New York" leaves the airport, date, and
service class unstated. \S\ref{sec:motivation} illustrates both concretely.
Without a formal semantic layer defining a task's required parameters, the
receiving agent must either guess, or fall into expensive, non-deterministic
clarification loops---computational cost that should have been eliminated at
the protocol level, not paid at inference time on every request.

\textbf{The Solution Framework:} Efficient multi-agent systems need protocols
that combine formal syntax with semantic knowledge---shared, machine-readable
schemas that let agents interpret content unambiguously and validate semantic
correctness before computation begins. This paper proposes two new
architectural layers: an \textbf{Agent Communication Layer (L8)} that
standardizes interaction structure, and an \textbf{Agent Semantic Layer (L9)}
that establishes shared meaning and \textbf{collective coordination} before
task execution. Together, they provide the foundation for scalable, distributed
agent collaboration---the essential infrastructure for an Internet of Agents
where computation scales through semantic coordination rather than context
window expansion.

\section{Background and Related Work}
High-level agent communication is not a new goal, but prior protocols
uniformly mistake syntactic agreement for semantic consensus.

\subsection{Classical Agent Communication Languages (ACLs)}
FIPA-ACL~\cite{fipa2002} moved beyond simple data transfer to formalize speech
acts~\cite{searle1969} (INFORM, REQUEST, PROPOSE), and its message structure
included an \texttt{:ontology} parameter, acknowledging a shared vocabulary was
necessary. It failed on two fronts: heavyweight formal ontologies (KIF, SL)
imposed overhead, and \texttt{:ontology} was merely a label---no mechanism
existed for aligning ontologies when agents lacked an identical, pre-shared
model~\cite{gruber1993}. It named the problem without solving it.

\subsection{Modern Agent Communication Protocols (A2A/MCP)}
\label{sec:motivation}
Modern successors -- A2A~\cite{a2a2024}, MCP~\cite{mcp2024}, and
NLIP~\cite{nlip2025} -- remedy FIPA-ACL's heavyweight ontologies with
lightweight JSON-style data-binding over modern web protocols. Like FIPA-ACL,
however, they excel at standardizing the syntactic
\textit{envelope} of communication while leaving semantic context unmanaged.
For example, a protocol might define a standard JSON structure for agent
messages:
\begin{verbatim}
JSON
{
  "protocol": "A2A/1.0",
  "performative": "REQUEST",
  "sender_id": "agent-travel-7",
  "receiver_id": "agent-booking-4",
  "content": {
    "task": "book_flight",
    "prompt": "I need a flight to New York for next Tuesday."
  }
}
\end{verbatim}
This message is syntactically perfect: a protocol-aware agent can parse it and
extract the task. But the protocol does not enforce or validate that both
agents share the same understanding of what "book\_flight" requires or what
"New York" means. The receiving agent has no guaranteed, protocol-driven way to
ground the natural-language string, so its LLM will be invoked on ambiguous
data. The agent may:
\begin{itemize}

\item \textbf{Guess:} Assume that "New York" means JFK and "next Tuesday" means the
immediate next Tuesday, potentially booking the wrong flight.
\item \textbf{Clarify:} Send a REQUEST-CLARIFY message back. This results in a
"prompt negotiation" loop, which is computationally expensive (each turn is a
full LLM inference) and non-deterministic.
\item \textbf{Fail:} Reject the request as too ambiguous.
\end{itemize}
The core problem: protocols formalize the request's structure but not its
context---the shared understanding of terms, goals, and constraints is left
unmanaged, producing brittle, unpredictable, inefficient systems. A separate
NLP/IR line of work studies clarification directly, learning \emph{when} and
\emph{what} to ask~\cite{aliannejadi2019asking,lee2023asking}; we instead
treat disambiguation as a protocol-level primitive triggered by
schema-validated grounding, at zero marginal cost when nothing needs asking.

\begin{figure}[t]
    \centering
    \resizebox{0.60\linewidth}{!}{%
    \begin{tikzpicture}[node distance=0cm,
        layer/.style={rectangle, draw=black, thick, minimum width=2.6cm, minimum height=0.6cm, align=center, font=\small},
        layer2/.style={rectangle, draw=black, thick, minimum width=4.8cm, minimum height=0.6cm, align=center, font=\small},
        newlayer/.style={rectangle, draw=blue!70!black, thick, fill=blue!10, minimum width=2.6cm, minimum height=0.6cm, align=center, font=\small},
        newlayer2/.style={rectangle, draw=blue!70!black, thick, fill=blue!10, minimum width=4.8cm, minimum height=0.6cm, align=center, font=\small},
        splitlayer/.style={rectangle, draw=black, thick, minimum width=2.4cm, minimum height=0.6cm, align=center, font=\small},
        splitnew/.style={rectangle, draw=blue!70!black, thick, fill=blue!10, minimum width=2.4cm, minimum height=0.6cm, align=center, font=\small}]

        % Traditional OSI Stack (left column)
        \node[layer] (app1) {Application (L7)};
        \node[layer, below=of app1] (pres1) {Presentation (L6)};
        \node[layer, below=of pres1] (sess1) {Session (L5)};
        \node[layer, below=of sess1] (trans1) {Transport (L4)};
        \node[layer, below=of trans1] (net1) {Network (L3)};
        \node[layer, below=of net1] (link1) {Data Link (L2)};
        \node[layer, below=of link1] (phy1) {Physical (L1)};
        \node[above=0.2cm of app1, font=\bfseries\small] {Traditional OSI};

        % Proposed Stack (right column)
        \node[splitlayer, anchor=north west, text width=2.2cm] (app_agent) at ($(app1.north east) + (0.8cm, 0)$) {Agentic \\ Application};

        \node[splitnew, below=of app_agent, anchor=north west] (sem3) at (app_agent.south west) {Agentic \\ Semantic (L9)};

        \node[splitnew, below=of sem3, anchor=north west] (comm3) at (sem3.south west) {Agentic\\Comm. (L8)};
        \node[splitlayer, anchor=north west, text depth=0.6cm] (app3) at ($(sem3.south west) + (2.4cm, 0)$) {App};

        \node[newlayer2, below=0cm of comm3, anchor=north west] (apptrans3) at (comm3.south west) {App Transport\\(L5-L7)};
        \node[layer2, below=of apptrans3, anchor=north west] (trans3) at (apptrans3.south west) {Transport (L4)};
        \node[layer2, below=of trans3, anchor=north west] (net3) at (trans3.south west) {Network (L3)};
        \node[layer2, below=of net3, anchor=north west] (link3) at (net3.south west) {Data Link (L2)};
        \node[layer2, below=of link3, anchor=north west] (phy3) at (link3.south west) {Physical (L1)};
        \node[above=0.2cm of app_agent, xshift=1cm, font=\bfseries\small, align=right] {Proposed Stack for Agents};

    \end{tikzpicture}%
    }
    \caption{Traditional OSI stack vs.\ our proposed stack: L8/L9 sit above
    HTTP/2/3, the Application Transport layer (L7).}
    \label{fig:stack_architecture}
\end{figure}

\section{A Layered Architecture for the Agentic Web}
\label{section:L9}
To solve this impasse, we propose a new architectural stack. We formally
recognize HTTP/2 and HTTP/3 as Layer 7, the \textbf{Application Transport
Layer}, whose FRAMES and STREAMS capabilities provide an ideal transport
foundation, and propose two new layers (L8 and L9) above it
(Figure~\ref{fig:stack_architecture}).

\subsection{Layer 8: The Agent Communication Layer}
The first new layer, L8, is the \textbf{Agent Communication Layer}: it ensures
agents can reliably exchange structured messages, understand intent, and
coordinate interaction flows, independently of semantic content, unifying the
best parts of A2A, MCP, NLIP, and FIPA-ACL into one standardized layer. L8
provides three components:
\begin{enumerate}
    \item \textbf{Message Structure:} A standard envelope --
    sender, receiver(s), message ID, performative -- already implemented via
    A2A~\cite{a2a2024} and MCP~\cite{mcp2024}'s structured messages, tasks, and
    artifacts.

    \item \textbf{Performatives:} The \textit{communicative
    intent} of a message, independent of its content. A2A's task requests,
    artifact exchanges, and status updates already map to distinct speech
    acts, which we generalize into a registry:
    \begin{itemize}
        \item \textbf{Transactional:} `REQUEST`, `AGREE`, `REFUSE`, `INFORM`
        \item \textbf{Negotiation:} `PROPOSE`, `ACCEPT`, `REJECT`, `COUNTER\_PROPOSE`
        \item \textbf{Information:} `QUERY`, `SUBSCRIBE`, `PUBLISH`
    \end{itemize}

    \item \textbf{Interaction Patterns:} Expected message sequence
    and coordination. A2A implements client-server today; SLIM~\cite{slim2025}
    adds publish-subscribe and group communication, together supporting:
    \begin{itemize}
        \item \textbf{Request-Reply:} A 1:1 `REQUEST` followed by `AGREE`/`REFUSE`.
        \item \textbf{Publish-Subscribe:} A 1:N `SUBSCRIBE`/`INFORM` topic pattern.
        \item \textbf{Aggregation:} An N:1 pattern collecting data from a
        population.
        \item \textbf{Collaboration Groups:} N:N coordination in secure
        groups.
    \end{itemize}
\end{enumerate}

This is what separates L8 from L9: L8 provides the ``how'' of communication,
while L9 (next) provides the ``what.''
\subsection{Layer 9: The Agent Semantic Layer (SL)}
Above L8 sits the \textbf{Agent Semantic Layer (L9)}: it establishes a
verifiable, shared meaning for the `content` payload of L8 messages and
facilitates \textit{coordination and consensus} across agent populations,
mediating semantic alignment \textit{before} the primary task is processed.

Unlike L8 -- where A2A~\cite{a2a2024}, MCP~\cite{mcp2024}, and
SLIM~\cite{slim2025} already provide substantial capabilities -- L9 is a
capability no current protocol offers. FIPA-ACL~\cite{fipa2002} acknowledged
the need for shared ontologies via its `:ontology` parameter, but no protocol
discovers, negotiates, and locks semantic contexts at the protocol level;
agents today rely on application-level convention or costly natural-language
negotiation.

L9 draws on distributed computing's shared-memory
coordination~\cite{chopra2025}: distributed processes share \textit{data
structures}, distributed agents must share \textit{meaning}. L9 lets agents
establish a "shared context" analogously, and, in population-scale systems
like the Ripple Effect Protocol, provides \textbf{consensus} primitives
(voting, quorum commit) so agents converge rather than diverge.

This context integrates into agent reasoning as both a \textit{validation
schema} the SL checks content against before it reaches the LLM, and via
injection into the LLM's own reasoning (system prompt or structured
retrieval).

\subsection{The Agent Semantic Layer (SL) in Detail}
The power of the L8/L9 split is separating communication structure (L8) from
meaning and collective state (L9). The SL negotiates and locks a
\textbf{Shared Context}: a formal, machine-readable definition (JSON Schema,
RDF/OWL, Protobufs) of the concepts, entities, tasks, and parameters relevant
to a domain, identified by a versioned URN (e.g.,
`urn:contexts:travel:v2.1`), that subsequent L8 messages adhere to.

\subsection{Functions of the SL}
The SL protocol consists of four primary phases:

\textbf{1. Context Discovery and Negotiation (The Semantic Handshake)}
Agents' SLs perform a handshake to find a compatible Shared Context:
\begin{itemize}
    \item \textbf{A (SL-HELLO):} "I support [`urn:contexts:travel:v2.1`,
    `urn:contexts:supplyChain:v1.0`]."
    \item \textbf{B (SL-SELECT):} "I also support `urn:contexts:supplyChain:v1.0`
    -- let's use that."
    \item \textbf{A (SL-LOCK):} "Locked. All `content` now validates against
    `urn:contexts:supplyChain:v1.0`."
\end{itemize}
This negotiation creates a cacheable, reusable "semantic session."

\textbf{2. Semantic Grounding}
\label{sec:l9-grounding}
Once locked, the SL grounds data---binding text, numbers, or vectors from the
agent's internal logic to the Shared Context's formal definitions---for both
transactional and coordinative patterns.

\textbf{3. Semantic Validation/Disambiguation}
The receiving SL validates incoming `content` against the locked context; an
undefined or multi-meaning concept triggers \textbf{semantic disambiguation}
-- the agents negotiate a shared meaning, which can then be added to the
context.

\textit{\textbf{Example: Booking a Flight}}
\begin{itemize}
    \item \textbf{Context:} `urn:contexts:travel:v2.1` defines `bookFlight`
    with required parameters `origin\_code`, `dest\_code`, `date`.
    \item \textbf{Agent A Logic:} "I need a flight to New York for next Tuesday
    from LAX."
    \item \textbf{Grounding:} The SL binds `origin\_code: "LAX"` and
    `date: "2025-11-04"`, but finds `dest\_code` ambiguous ("New York").
    \item \textbf{SL-driven Clarification:} Instead of guessing, the SL uses
    L8 to send a formal query: `QUERY` with content
    `{"concept": "ambiguous\_parameter", "parameter": "dest\_code", "value":
    "New York"}`.
\end{itemize}

\textbf{4. Coordination and Consensus}
In population-scale systems, grounding is insufficient; agents must agree. L9
provides \emph{Consensus Primitives}~\cite{chopra2025}: a proposed state
change "ripples" until a quorum commits, preventing stale-belief divergence.
We treat this as architectural, not an evaluated claim; \S\ref{sec:evaluation}
covers grounding and disambiguation only.

\section{Evaluation}
\label{sec:evaluation}

We evaluate whether the L8/L9 protocol's grounding and disambiguation
round-trip (\S\ref{section:L9}) measurably improve accuracy and cost relative
to two ad-hoc alternatives, and run an ablation (\S\ref{sec:ablation})
isolating whether the gain comes from the schema itself or merely an extra
exchange.

\paragraph{Datasets.} Both datasets' annotations supply a ground-truth
interpretation, letting us build a deterministic oracle responder without
hand-labeling. \textbf{AmbigNQ}~\citep{min2020ambigqa} is open-domain QA with
genuinely ambiguous questions (multiple sub-question interpretations) and
unambiguous ones, playing the JFK/LGA/EWR role above. \textbf{SGD}~\citep{rastogi2020schema}
is task-oriented dialogue whose \texttt{Flights\_1} schema gives one
categorical slot (\texttt{airlines}, 8 candidates) plus three free-text slots
per booking intent; an item is \emph{ambiguous} if the categorical slot is
unresolved once its free-text slots resolve.

\paragraph{Pipelines and models.} Only \textsc{SemanticStack} has the schema.
\textbf{Guess} issues one blind call, no chance to ask (\S\ref{sec:motivation}).
\textbf{Clarify} lets the agent decide whether to ask, in natural language,
first. \textsc{SemanticStack} makes one grounding call deciding, per field,
\emph{grounded}/\emph{ambiguous}/\emph{missing}, triggering one QUERY/INFORM
round-trip per unresolved field. We run every pipeline on two LLMs of very
different capability,
\textbf{Claude Sonnet}~\citep{anthropic2026claudesonnet46} and
\textbf{Gemma}~\citep{gemmateam2026gemma4}.

\paragraph{Metrics.} A prediction is correct if it or the gold answer is a
normalized substring of the other (for SGD, per slot, plus all-slots-correct).
We report average LLM calls, tokens, and round-trips per item; any outright
failure (malformed response or 240s timeout) scores as incorrect.

\subsection{Main Results}
\label{sec:main-results}

Table~\ref{tab:main-results} reports accuracy and efficiency on the full
evaluation sets (AmbigNQ: 1{,}172+830 items; SGD \texttt{Flights\_1}:
57+273), split by whether the item is genuinely ambiguous.

\begin{table}[t]
\centering
\footnotesize
\setlength{\tabcolsep}{4pt}
\renewcommand{\arraystretch}{0.9}
\begin{tabular}{lll cc cc}
\toprule
& & & \multicolumn{2}{c}{Ambiguous} & \multicolumn{2}{c}{Unambiguous} \\
\cmidrule(lr){4-5} \cmidrule(lr){6-7}
Dataset & Model & Pipeline & Acc.\ (\%) & Calls & Acc.\ (\%) & Calls \\
\midrule
\multirow{6}{*}{AmbigNQ}
  & \multirow{3}{*}{Sonnet} & Guess              & 35          & 1.00 & 69          & 1.00 \\
  & & Clarify                & 44          & 1.10 & \textbf{73} & 1.01 \\
  & & \textsc{SemanticStack} & \textbf{52} & 2.00 & 69          & 1.00 \\
\cmidrule(lr){2-7}
  & \multirow{3}{*}{Gemma}  & Guess              & 25          & 1.00 & 50 & 1.00 \\
  & & Clarify                & 28          & 1.31 & 50 & 1.24 \\
  & & \textsc{SemanticStack} & \textbf{38} & 2.00 & 50 & 1.00 \\
\midrule
\multirow{6}{*}{SGD \texttt{Flights\_1}}
  & \multirow{3}{*}{Sonnet} & Guess              & 0           & 1.00 & 23          & 1.00 \\
  & & Clarify                & 67          & 2.93 & 33          & 1.96 \\
  & & \textsc{SemanticStack} & \textbf{72} & 1.00 & \textbf{66} & 1.00 \\
\cmidrule(lr){2-7}
  & \multirow{3}{*}{Gemma}  & Guess              & 16          & 1.00 & 29          & 1.00 \\
  & & Clarify                & 75          & 3.00 & 52          & 2.07 \\
  & & \textsc{SemanticStack} & \textbf{88} & 1.00 & \textbf{86} & 1.00 \\
\bottomrule
\end{tabular}
\caption{\textsc{SemanticStack} improves ambiguous-item accuracy over blind
Guessing on every dataset/model pair (+13--17 points on AmbigNQ;
0--16\%$\to$72--88\% on SGD) while matching Guess's cost of one call whenever
the item is unambiguous.}
\label{tab:main-results}
\end{table}

Three patterns hold across both datasets and models: \textsc{SemanticStack}
always beats Guessing on ambiguous items (+13--17 on AmbigNQ;
0--16\%$\to$72--88\% on SGD); it costs \emph{nothing} extra when unneeded,
matching Guess's 1.00 calls unlike Clarify's occasional unnecessary asks
(1.01--2.07); and it holds even for Gemma, substantially weaker (3--7\%
AmbigNQ failure rate, counted against it), with an even larger SGD gain.

\subsection{Ablation: Isolating the Role of the Schema}
\label{sec:ablation}

\textsc{SemanticStack}'s round-trip alone is not evidence that
schema-grounded disambiguation matters: an agent might do as well asking
\emph{any} clarifying question, since our oracle responder always answers
correctly once asked. To isolate the schema's contribution,
\textsc{SemanticStack-NoSchema} runs the identical protocol and round-trip,
but grounding never sees the domain's candidate values, so it judges
ambiguity/completeness from the bare utterance alone. Table~\ref{tab:ablation}
reports Claude Sonnet results (full SGD set; a 300-item AmbigNQ subsample
whose \textsc{SemanticStack} row tracks Table~\ref{tab:main-results} within 1
point; Guess/Clarify shift 2--3 points at this size).

\begin{table}[t]
\centering
\footnotesize
\setlength{\tabcolsep}{4pt}
\renewcommand{\arraystretch}{0.9}
\begin{tabular}{ll cc cc}
\toprule
& & \multicolumn{2}{c}{Ambiguous} & \multicolumn{2}{c}{Unambiguous} \\
\cmidrule(lr){3-4} \cmidrule(lr){5-6}
Dataset & Pipeline & Acc.\ (\%) & Calls & Acc.\ (\%) & Calls \\
\midrule
\multirow{4}{*}{AmbigNQ}
  & Guess                          & 33          & 1.00 & 71 & 1.00 \\
  & Clarify                        & 41          & 1.08 & 71 & 1.00 \\
  & \textsc{SemanticStack}         & \textbf{51} & 2.00 & 71 & 1.00 \\
  & \textsc{\phantom{x}--NoSchema} & 35          & 2.00 & 71 & 2.00 \\
\midrule
\multirow{4}{*}{SGD \texttt{Flights\_1}}
  & Guess                          & 0           & 1.00 & 23          & 1.00 \\
  & Clarify                        & 68          & 2.93 & 34          & 1.97 \\
  & \textsc{SemanticStack}         & 72          & 1.00 & 66          & 1.00 \\
  & \textsc{\phantom{x}--NoSchema} & \textbf{72} & 1.00 & \textbf{67} & 1.00 \\
\bottomrule
\end{tabular}
\caption{Ablation (Sonnet). Removing the candidate list collapses most of
\textsc{SemanticStack}'s AmbigNQ advantage (51\%$\to$35\%, next to Guess's
33\%) but leaves SGD unchanged (72\%$\to$72\%; 66\%$\to$67\%), and costs
NoSchema the zero-overhead property on unambiguous items (1.00$\to$2.00
calls).}
\label{tab:ablation}
\end{table}

The two datasets diverge informatively: the schema is load-bearing on AmbigNQ
but not on SGD. An open-domain question's valid readings form an unbounded,
domain-specific space no general-purpose judgment can enumerate; but noticing
an utterance never states an airline preference is a generic
missing-information judgment, needing no knowledge of the eight named
candidates. The schema matters precisely when detection itself is
domain-specific; when it reduces to noticing an omission, the L9 protocol's
structural guarantee alone carries the result, and the schema's remaining
contribution is cost (Table~\ref{tab:main-results} vs.\ Clarify's 2--3 calls)
rather than accuracy.

\paragraph{Limitations.} Our disambiguation responder is a deterministic
oracle for both \textsc{Clarify} and \textsc{SemanticStack}; a real asker
would be noisier, and results are single-run with no variance/significance
testing. SGD coverage is one service and one categorical-slot pattern; the
substring-match metric is unvalidated against human judgment; and security
(injection, poisoning) is deliberately out of scope, not merely unevaluated.

\section{Conclusion}
As autonomous agents become significant network actors, the protocol stack
may need to evolve again, as it has before (IP, TCP, HTTP/2/3). This paper
proposes a two-layer extension. \textbf{L8} formalizes structural
capabilities already emerging in A2A~\cite{a2a2024}, MCP~\cite{mcp2024}, and
SLIM~\cite{slim2025}. \textbf{L9} introduces protocol-level semantic
negotiation and \textbf{collective consensus}---a gap left to informal
convention today. Our evaluation shows schema-grounded disambiguation
improves accuracy over ad-hoc negotiation at no added cost when unneeded
(\S\ref{sec:evaluation} discusses remaining limitations).

%\section{Acknowledgements}
%\input{acknowledgements}

\bibliographystyle{ACM-Reference-Format}
\bibliography{main}
\end{document}